# POLARIZATION THEORY OF MOTIVATIONS, EMOTIONS AND ATTENTION

S.E. Murik
Irkutsk State University
Department of Physiology and Psychophysiology
Irkutsk, Russia
sergey_murik@yahoo.com

A new theory of motivations, emotions and attention is suggested, considering them as functions of sensory systems. The theory connects neurophysiological mechanisms of mental phenomena with the change of metabolic and functional state of perceptive neurons, which is reflected in the degree of polarization of a cell membrane.

One of the fundamental challenges of modern natural sciences is the problem of relating the physiological background of human and animals to - mental activity. The state of this issue can hardly be considered to be satisfactory. Over the last century biological sciences such as genetics, molecular biology, biophysics and others have achieved obvious success. Nevertheless, physiology of mentality has not been able to solve the key problem, facing it. For more than a hundred years we do not know any real mechanisms of mental phenomena. All our knowledge is nothing more than a huge amount of accumulated experimental data, which are hardly put into schemes, which with caution can be named a real arrangement and mechanisms of the brain function.

«A thought is the first two thirds of the mental reflex» – I.M. Sechenov said (1866, p. 99). Today we can add nothing more to these words about mechanisms of thinking.

One of the first natural scientific ideas on mechanisms of emotions and motivations belong to W. James (1884) and W.B. Cannon (1929). The essence of the ideas is that emotions and motivations are the result of perception of irritants by the brain. At present the scientists are 100 % sure that actually these phenomena are born in the nervous system, but what their neuronal mechanism is we still do not know definitely.

At present the mechanism of motivation is considered in the context of existence of specific and nonspecific motivatiogenous centers and «central motivational excitation» (Stellar, 1954; Dell, 1958; Morgan, 1959; Olds, Olds, 1965; Sudakov, 1971; Simonov, 1987). Despite a long history of physiology of motivations problem, neither the nature of motivational excitation, nor neurophysiological mechanisms of motivational states are clear enough.

The state of the problem of physiological basis of emotions is unsatisfactory. There is one statement that is actual today like twenty years ago which says that at present there is no uniform standard scientific theory of emotions and also exact data of how and in what centers these emotions arise and what their nervous substrate is (Schmidt, Thews, 1983).

The modern approach to the problem of motivations and emotions is peculiar by consideration of their mechanism outside of sensory systems and accordingly outside of processes of perception. This approach is defined by domination of cybernetic ideas in physiology of mental activity (Winer, 1948). The essence of the approach is in allocation of independent structural-functional elements in the brain, their ordering and co-subjecting by mechanistic principles. Research of the brain in light of the given outlook presupposes exarticulation of systems of management and control in the brain and their introduction in mechanisms of the brain function.

Present-day physiology of motivations and emotions not only excludes their mechanisms out of an act of perception but also allocates separate phenomena (minorant functions) with independent brain substrata in these mechanisms (Simonov, 1981; 1987). Finally the brain becomes indefinitely complex mechanical nested doll in which this or that system consists of several elements which in their turn consists of their own elements and so on and so forth. To establish character of interrelations and relations in such a system between great number of elements of different levels becomes indefinitely complex task. That is probably why by the present time there is no advance in knowledge of mechanisms of mental activity.

In our opinion, the way out is to give up the system (Bertalanfy, 1969) and cybernetic (Winer, 1948) principles in solution of problem of structural-functional arrangement of the brain and to consider neurophysiological mechanism of motivations, emotions and attention in connection with the change of functional state (FS) of afferent (perceptive) neurons.

Nowadays the term FS is widely used in physiology to study different levels of human and animal organism, but at the same time there may be an absolutely different interpretation of this term. In neurophysiology, speaking of FS of neurons such phenomena as excitation, inhibition and rest are implied (Vvedenskiy, 1901; Golikov, 1950; Golikov, Kopilov, 1985; Movchan, 1985).

Sensu stricto, functional state means the state of function of realization. Speaking of FS it is necessary to describe in what state the realization of this or that function is: in good or bad, i.e. to give qualitative description of the function.

Long since the idea of FS as a qualitative description of activity in contrast to neuronal level is used in studies of the whole organism. From this point of view efficacy of behavior is an integral parameter of the organism's FS (Medvedev, 1970; Leonova, 1984). Apparently, the efficacy of behavior depends on functional states of lower level element composing the organism. However, even if the concept of FS is used on the level of systems of organs as qualitative description of their functioning, it practically never happens on the cellular level.

Based on the concept of FS as qualitative description of functioning, functional state of a neuron can be defined as an index of ability of a cell to perform functions peculiar to it at present point in time.

The main, if not the only function of neurons established with some degree of certainty is generation of nerve impulses or action potentials (AP). Correspondingly, measure of neurons activity efficacy should be an index of efficacy of this function, i.e. productivity of impulse activity.

In physiology this characteristic of excitable formation is called functional mobility or lability (Vvedenskiy, 1901). Considering lability as a measure of effectiveness of neuronal activity in physiology is recently very rarely met. However, exactly lability reflects per se FS of a neuron.

The analysis of literature data shows that the efficacy of neuron activity (functional ability) changes at changing level of membranous potential (MP). Depolarization is accompanied by reduction, and hyperpolarization – by increase of lability of excitable formation (Sologub, 1985), according to that FS of neuron changes: at hyperpolarization it improves, at depolarization – deteriorates.

To the fact that stationary depolarization of MP is unfavorable not only from the functional, but also from the metabolic point of view testify great amount of experimental data.

In the first place, it is long known (Shapot, Gromova, 1954) that appearance of stationary depolarization of neuronal membrane can happen only if mechanisms reconstructing ionic homeostasis after generation of impulses, can not fulfill their task. More often, it happens at energy hunger of a cell. In other words, development of neuronal depolarization in the process of activity reflects the appearance of macroergic compounds deficit.

In the second place, it has been shown that stationary depolarization starts up a whole cascade of pathogenic processes, which in the end cause the cells' death. Thus, in the beginning potential-depending $Ca^{2+}$-canals open, $Ca^{2+}$ entering a cell stimulates excretion of exciting amino acids (glutamate, aspartate, etc.) and activation of potential-independing $Ca^{2+}$, $Na^+$, $K^+$ and $CL^-$ canals. Accumulation of intracellular calcium provokes cascades of biochemical reactions of free radicals and lipid peroxidation. Glutamate-dependent increase of intracellular $Ca^{++}$ reduces MP of mitochondria (Isaev et al., 1994). As a consequence: swelling of mitochondria, damage of external membrane and exit of mitochondria out of intermembranous space into cytosol of albumen, causing apoptosis.

There is no escape from admitting that stationary depolarization of cellular membranous potential in intervals between action potentials is not an indifferent state from the point of view of quality of metabolism, but it testifies to predominance of catabolic processes and to the lag of anabolic processes behind vital need of a cell, i.e. it testifies to deterioration of vital phenomena in a cell.

As mentioned before, lability of a neuron increases with hyperpolarization of membranous potential (Sologub, 1985). Literature analysis (Andersen, Eccles, 1962; Batuev, 1970; Skrebitskiy, 1977 et al.) also shows that in connection with processing of information from internal and external medium of the organism, activation of neurons is preceded exactly by hyperpolarizational deviation of MP, and generation of AP looks like posthyperpolarization recoil. Consequently, hyperpolarization can be considered as reflection of operational FS of nervous cells. Since hyperpolarization deviation of MP is obligatory primary reaction of nervous cells on any impact (Vvedenskiy, 1901; Vasiliyev, 1925), the state forming at that, in all probability, reflects mobilization of adaptive intracellular mechanisms. The fact that during hyperpolarization adaptive reserves of cells are mobilized is proved by experimental data, which testify to the increase of cells' resistance in this period (Nasonov, 1959). It is also known that hyperpolarizating preparations very often have protective effect in conditions unfavorable for neurons life (Kulinskiy et al., 1994; Sufianova et al., 2003).

Nowadays we can claim with confidence that FS of a neuron (lability) depends of ability of the structure to restore its energetic and plastic recourses, necessary for conducting of valuable reaction. In other words, FS as characteristic of efficacy of neurons performance of their functions associates with cellular metabolic processes. Neurons performance of functions influences upon their metabolic state, the shifts of which, in their turn, determine functional abilities of cells. That is why FS shifts are always interrelated with shifts of metabolic state, and FS indexes also characterize metabolic state of excitable formation. Therefore, changing of lability may be considered as an indicator of shifts in functional and metabolic state of excitable tissue. Reduction in lability testifies to bad, and increases – to good functional and metabolic state of considered nervous formation.

Consequently, literature data analysis testifies to that integral index of metabolic and functional state of brain neurons is the level of membranous potential. Totality of biochemical and electrophysiological data indicates that stationary depolarization of cellular MP reflects development of unfavorable functional and metabolic state in them, that is why it should be undesirable both to separate cells and to the whole organism. Development of unfavorable FS in specified populations of neurons, in this case, may be the basis for nervous system's revealing stimuli and factors having negative biological meaning for the organism; and also it may be the basis for organization of adaptive behavior on this background. In other words, biological significance of a stimulus may be valued according to functional state of neurons perceiving it.

Development of unfavorable FS, like stationary depolarization can be neurophysiological basis of motivated states and of subjective experience in the form of negative emotions. Then, elimination of mentions state in the organism, i.e. repolarization and hyperpolarization of neurons by humoral and behavioral reactions will be a reward and mechanism of positive emotions.

As our experiments on rats show, at motivated states and negative emotions, which accompany such states, in the nervous tissue of the brain depolarization processes start to occur similar to those which are observed at superficial ischemia (Murik et al., 2003).

The research of polarization processes in the nervous system becomes possible with the use of registration of so called steady potential or direct current potential (Marczynski, 1993; Kohling et al., 1996). At that negative shift of steady potential reflects depolarization of MP of nervous tissue cells. At the same time, positive reverse process reflects repolarization or hyperpolarization.

As seen in Fig. 1, relatively weak brain ischemia, connected with partial limitation of brain blood current ("Ischemia-1"), was accompanied by an increase of EEG rhythm power and by a small negative shift of steady potential. On average lowering of potential in all sampling (n = 24) made 1222.51±290.1 mkV. Even greater limitation of brain blood current by means of additional exclusion of median brain artery ("Ischemia-2") brought to depression of EEG on the background of significant negative shift of steady potential.

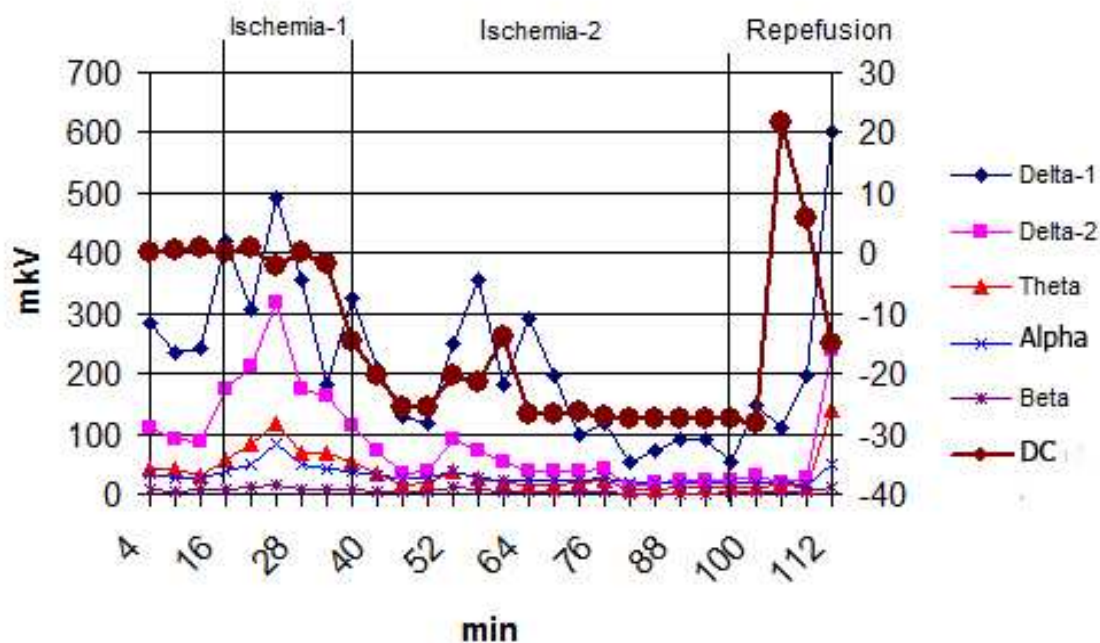

Fig. 1. Change of steady potential and EEG in parietal cortex of the left cerebral hemisphere in the rat at modeling of brain ischemia by different means (Murik et al., 2003). Designations: "Ischemia-1" – binding up of common carotids; "Ischemia-2" – insertion of occluder in median brain artery (MBA) of the left hemisphere; "Reperfusion" – withdrawal of the occluder from MBA. EEG rhythm amplitude (mkV) is shown on the left scale of Y axis, and the DC potential level (mV), in the right one. DC potential values in the period, preceding "Ischemia-1", are taken as a zero.

Fig. 2 shows the change of the same indexes after injection of narcotic into abdominal cavity of rats. It is seen that at once after injection of the preparation negative deviation of steady potential level and increasing of EEG power were observed. Negative shift of the potential made on average 638.8±124.1 mkV.

Thus, negative emotional excitement, developing apparently after fixation of them in hands, puncturing of coverlet with a needle and injection of preparation, was accompanied by negative shift of steady potential and increasing of EEG power. As a whole it corresponded with electrophysiological changes, observed during "Ischemia-1" (Fig. 1) and testifies to similarity of phenomena occurring in both cases on neuronal level. In either case, the state of depolarization exaltation of neurons excitability and strengthening of nervous tissue metabolism evidently occurred according catelectronic type.

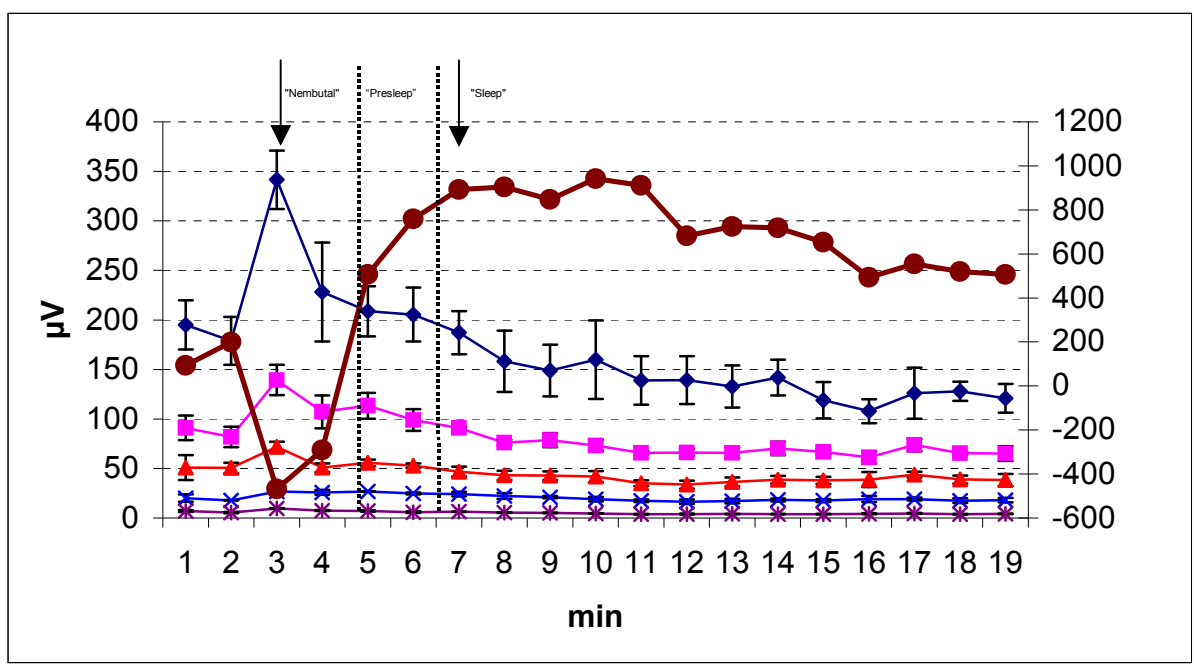

**Fig. 2.** Change of steady potential level and amplitude of EEG rhythms after intraperitoneal injection of Nembutal (Murik, Shapkin, 2004). EEG rhythm amplitude (mkV) is shown on the left scale of Y axis, and the DC potential level (mkV), in the right one.

As Nembutal was absorbed into blood, positive deviation of steady potential level with preserved increased power of EEG (on the Fig. 2 this period is marked as "Presleep"). The character of steady potential level change (namely its positivation) testifies to development at that re- and hyperpolatization processes. Increased amplitude of rhythms indicates that hyperpolarization inhibition was absent at that time. Comparison of characteristics of steady potential level and EEG allows viewing the development of functional state, corresponding to anode exaltation and strengthened metabolism of nervous tissue, in neurons during "Presleep" period. Finally, beginning and development of narcotic sleep was accompanied by even greater positive shift of steady potential, combined with oppression of EEG power. This functional state reflects, evidently, deepening of hyperpolarization of cells and beginning of hyperpolarization inhibition together with reduction of nervous tissue metabolism.

It is clear that on the whole the character of polarization processes after injection of narcotic preparation was opposite to that one observed at ischemia. It indicates that at the same time another metabolic and functional state of brain neurons developed: ischemia – a model of formation of unfavorable, and Nembutal narcosis – of good functional and metabolic state of the nervous tissue.

We observed development of depolarization processes reflecting formation of unfavorable metabolic and functional state in animals in conditions of hunger. When in structure related to food center (lateral hypothalamic nucleus) food motivation was formed, depolarization made 460 mkV (Murik, 2002). Process of food consumption and the state of saturation were accompanied by repolarization processes in the mentioned structure.

As our experiments showed (Fig. 1–2), activation of neurons is possible on the background of both depolarization and hyperpolarization processes, i.e. at their both bad or good metabolic and FS. Generation of action potentials following the pattern of posthyperpolarization recoil is more effective mechanism of excitement than any other one from metabolic point of view. And it correspondingly can be considered as the optimum (operational) mechanism in the nervous system. Hyperpolarization wave, preceding excitement, evidently, mobilizes homeostatic adaptive mechanisms. In this state a cell is able to generate greater amount of action potentials and to do it easier than in any other state (Sologub, 1985). Hyperpolarization deviation of membranous potential increases functional abilities of the cell and thereby its adaptive ability. Thus, hyperpolarization inhibition is not an inhibition in its essence, but the state of increased resistance and complete readiness of functional and metabolic recourses of the cell (Sologub, 1985). If we consider it from the qualitative point of view – it is an excellent functional state, which is favorable for fulfillment of functions peculiar to a nervous cell, in the first turn – generation of action potentials.

Thus, equalization, from functional point of view, of excitement and inhibition, considering them as equivalent operational acts unjustified from biological positions. The state of vital processes at excitement and inhibition (both hyperpolarization and depolarization inhibition) differs greatly. That is why the processes of reacting to these or those factors on the level of individual cells (excitement and inhibition), most likely should be considered as different adaptive acts with corresponding characteristics of strain of intracellular adaptation mechanisms and qualitative changes of metabolism in them.

The experiments presented show the negative emotional excitement and the activation of orientative-explorative motivation to be followed by the formation in the neocortex of the unfavorable metabolic and functional state similar to a weak ischemia. According to bioelectrical pattern in both cases the nervous activation is observed followed by development of depolarization processes. The state of narcotic intoxication and apparent experiencing of positive emotions are combined with excitation at the background of hyperpolarization processes in the nervous system. High depolarization of brain neurons and following development of unfavorable metabolic and functional state were observed in hungry rats. As the rats were fed, repolarization processes were evolved in the central nervous system of the animals.

Hence, motivation states and experiencing negative emotions result from development of unfavorable metabolic and functional state of neurons appeared in depolarization of membrane cells. Positive emotions are caused by restoration of effective metabolic and functional state and repolarization of the brain cells as well as the development of excellent metabolic and functional state expressed in hyperpolarization phenomena.

The totality of accumulated by the present moment physiological data without controversy indicated that stationary depolarization of membranous potential reflects the development of unfavorable metabolic state, and therefore could not be desirable both for individual cells and for the whole organism. In this case, the character of polarization shifts should designate biological significance of irritants and changes of medium factors. Development of unfavorable metabolic state would testify to negative biological features of operating agents, whereas good metabolic state developing under the influence of these or those factors would testify to positive ones.

Development of unfavorable metabolic and functional state in neuronal systems, perceiving irritants and changes of medium states, can't help affecting subjective world of the individual in the form of negative emotional experience. That is why the totality of literature data, available and obtained by us, with much confidence makes possible to say that change of metabolic and correspondingly of functional state of perceptive neurons should be considered as possible neurophysiological basis of emotions. The same neurons' mobilization of system mechanisms of adaptation in the form of activation of behavioral engrams is as well the basis for purposeful (motivated) behavior. Consequently, neurophysiological mechanism of motivations and emotions evidently merges with sensory processes and consists in change of metabolic and functional state of perceptive neurons.

The approach we develop to the content of FS idea on the cellular level gives possibility to revise cardinally mechanism of complete behavioral act and components comprising it. In particular, neurophysiological basis of attention is evidently very tightly connected with functional state of perceiving neurons. Formation of engrams, including neurons with unfavorable functional state, in afferent systems under irritants affect will evidently be "a signal" for perceiving system to organize "reference behavior" directed at elimination (reduction) of the given state. In other words, from the point of view of the developed conception, attention is variety of motivation. Hence, functional state of neurons of nervous model of stimulus may be a universal mechanism, underlying many psycho-physiologic phenomena, including attention.

Until the present time, brain mechanisms of the mentioned mental phenomena were considered out of analyzing systems; and in the end it predetermined absence of real advance in comprehension of them. The majority of known theories in compliance with cybernetic (mechanistic) approach marked out independent brain systems for these mechanisms. All modern and popular theories position brain substrate of emotions and motivations in so called limbic system (Olds, Olds, 1965; Anokhin, 1964; Gelgorn, Loufborrow, 1966; Simonov, 1981, 1987). Cybernetic idea of existing in the brain independent neuronal system of emotions and motivations does not find any actual substantiation. Critical analysis of structural-functional arrangement of limbic cortex and other parts of the brain associated with it and comprising limbic system (MacLean, 1955, 1970), does not give any grounds to consider them as central nervous substrate of subjective experience. Existing morpho-functional data (Vasilevskaya, 1971; Zambrzhitskiy, 1972; Beller, 1983 et al.) undoubtedly show only that limbic system includes nervous elements of visceral analyzer and efferent systems, regulating activity of internal organs and the state of internal medium of the organism. In other words, if limbic system has any attitude to emotions, that is only to experience of visceral stimuli or as nervous substrate of vegetative and partially somatic congenital efferent reactions accompanying emotions. In that case, the approach implying availability of independent nervous center in emotional (subjective) experiences remains unsubstantiated.

The conception that we work out of functional state of neurons and its role in mechanisms of mental activity makes possible to speak about necessity of change of mechanistic approach to the problem of structural-functional arrangement of the brain dominating at present. Inasmuch as current state of a neuron in its essence is a function derived from activation of its adaptation mechanisms, the whole nervous system and its activity can be regarded as totality of cells adapting to irritants and changes of medium conditions which through the lowering of strain of its own adaptation mechanisms promotes adaptation of the whole multicellular colony of the organism. The nervous system's promotion of adaptation of the organism to the surroundings through mobilization of intrasystem mechanisms is realized through generation of action potentials by neurons.

Considering of functional state of brain neurons as an index of efficacy of adaptation mechanisms give a new view on physiological mechanism of many mental phenomena and their manifestation in norm and pathology, and also it gives a new view on nature of human and animal organism as a whole. So functional role of nervous system consisting in promotion efficient organism adaptation in the view of developed approach is realized through activation of neuronal adaptation mechanisms. The whole organism adapts through the neuronal adaptation: when cells feel well, it means that the whole cellular colony, the whole organism feel well.

## Definitions of motivations, emotions and attention from the point of view of polarization theory

**Motivation** is the process of mobilization of brain neurons, being in unfavorable functional (metabolic) state and manifesting in stationary depolarization of membranous potential, mechanism of restoration of their good functional state.

**Emotion** is a mental phenomenon of a subject's experience of changes of functional (metabolic) state of neurons of afferent (sensory) brain systems, appearing at influencing of irritants from external and internal medium of an organism.

Deterioration of functional (metabolic) state of brain neurons consisting in their depolarization is subjectively experienced in the form of negative emotion. In other words, negative emotions are subjective experience of formation of motivated state.

Improvement of functional (metabolic) state of brain neurons consisting in repolarization of membranous potential is subjectively experienced as a positive emotion.

Excellent functional (metabolic) state of brain neurons reflecting increasing of cellular resistance and manifesting in hyperpolarization of cells is subjectively experienced in the form of strong positive emotions.

**Attention** is a variety of motivation. It represents formation of unfavorable functional (metabolic) state of sensory neurons at effect of irritants predominantly from external medium. It differs from motivation only in degree of unfavorable state and in volume of nervous tissue it seized. Attention is connected with relatively not deep shifts of functional state in small volumes of nervous tissue, approaching nervous models of stimuli. Motivations seize by unfavorable functional state the whole sensory systems.

References

Andersen, P., Eccles, J. (1962). Inhibitory phasing of neuronal discharge. *Nature*, **196**, 645–647.

Anokhin, P.K. (1964). Emotions. *The Big Medical Encyclopedia*, 2nd ed.. **35**, p. 339-357.

Batuev, A.S. (1970). *Functions of movement analyzer*. Leningrad: Publishing House of Leningrad State University.

Beller, N.N. (1983). *The arrangement and mechanisms of central efferent influences upon visceral functions*. Leningrad: Nauka.

Bertalanfy, L.F. (1969). General theory of systems // *System researches*. Moscow, p. 30–34.

Cannon, W.B. (1929). Hunger and thirst. In C. Murchison (Ed.), *The foundations of experimental psychology* (pp. 434–448), Worcester, Mass., Clark University Press.

Dell, P.C. (1958). Some basic mechanisms of the translation of bodily needs into behavior. In G.E.W.Wolstenholme & C.M.O'Conner (Eds.) *Symposium on the neurobiological basis of behavior*. Boston: Little, Brown, p.187.

Gellhorn, E., Loofbourrow, D. (1966). *Emotions and emotional disorders*. Moscow: Mir.

Golikov, N.V. (1950). *Physiological lability and its change at main nervous processes*. Leningrad:Nauka.

Golikov, N.V., Kopilov, A.G. (1985). Researches of L.L. Vasiliyev about main nervous processes and their place in modern physiology. In N.P.Movchan (Ed.), *Physiological mechanisms of main nervous processes*: *Proceeding of Leningrad society of natural scientists*. Leningrad, **75**, Issue 5, p. 15–23.

Isayev, N.K., Zorov, D.B., Lizhin, A.A. et al. (1994). Glutamate causes reduction of membranous potential of mitochondria in cultivated cells-granules of cerebellum. *Bulletin of experimental biology and medicine*, № 2, 208.

James, W. (1884).What is emotion. *Mind*, № 4, 188.

Kohling R., Schmidinger A., Hülsmann S., Vanhatalo, S., Lücke, A., Straub, H., Oppel, F., Greiner, C., Moskopp, D., Wassman, H. (1996). Anoxic terminal negative DC-shift in human neocortical slices in vitro // Brain Research, Nov 25, **741** (1-2), 174–179.

Kulinskiy, V.I., Usov, L.A. Sufianova, G.Z., Sufianov, A.A. (1994). Protective effect of intercerebrovenrticular injection of A-Agonists at complete ischemia of the cerebrum. *Bullitin of experimental biology and medicine*, № 6, 622–624.

Leonova, A.B. (1984). *Psychodiagnostics of functional states of a man*. Moscow: Publishing House of Moscow State University.

MacLean, P. (1955). The limbic system (visceral brain) in relation to central gray and reticulum of the brain stem. Psychosom. Med., **17**, p-p 355-366.

MacLean, P. (1970). The limbic brain in relation to the pshychoses. In P.Black (Ed.), *Physiological correlates of emotion* (pp.129-146). New York, London:Acad. Press.

Marczynski, T.J. (1993). Neurochemical interpretation of cortical slow potentials as they relate to cognitive processes and a parsimonious model of mammalian brain. In: McCallum W.C., Curry S.H., (Ed.), *Slow potential changes in the human brain* (pp. 253-275). New York:Plenum Press.

Medvedev, V.I. (1970). Functional states of an operator. *Ergonomics: Principles and recommendations*. Moscow, Issue 1, 127–160.

Morgan, C.T. (1959). Physiological theory of drive in Psychology: A Study of a Science (Sensory, perceptual and physiological formulations). Koch S. ed., **1**, New York, McGraw, 645.

Movchan, N.P. (1985). Researches of L.L. Vasilyev – a new step in development of the doctrine of N.E. Vvedenskiy about parabiosis. In N.P.Movchan (Ed.), *Physiological mechanisms of main nervous processes*: *Proceeding of Leningrad society of natural scientists*. Leningrad, **75**, Issue 5, p. 5–15.

Murik, S.E. (2002). Approach to the study of neurophysiological mechanism of food motivation. *International Journal of Neuroscience*, **112**, 1059-1072.

Murik, S.E., Shapkin, A.G. (2004). Simultaneous recording of the EEG and direct current (DC) potential makes it possible to asses the functional and metabolic state of the nervous tissue. *International Journal of Neuroscience*, **114**, 921-941.

Murik, S.E., Sufianov, A.A., Sufianova, G.Z., Shapkin, A.G. (2003). Experimental data on electrophysiological correlates of brain ischemia of different severity. *Bulletin of Eastern-Siberian Scientific Center SB RAMS*, № 1, 148–154.

Nasonov, D.N. (1959). *Local reaction of protoplasm and distribution of excitement*. Leningrad.

Olds, J., Olds, M.E. (1965). Drives, rewards, and the brain. In New directions in physiology, vol. 2, New York, Holt, 327-410.

Schmidt, R., Thews, G. (Eds.) *Human Physiology*, vol.1, Springer Verlag, 1983.

Sechenov, I.M. Reflexes of the cerebrum (1866). *Selected works*. Moscow, 1953.

Shapot, V.S., Gromova, K.G. (1954). Energetic exchange of the cerebrum and the problem of hypoxic states // Biochemistry of the nervous system. Kiev, p. 139–150.

Simonov, P.V. (1981). *Emotional brain*. Moscow: Nauka.
Simonov, P.V. (1987). *Motivated brain*. Moscow: Nauka.
Skrebitskiy, V.G. (1977). *Regulation of conduction of excitement in visual analyzer*. Moscow: Medicina.
Sologub, M.I. (1985). Functional characteristics of cells at their hyperpolarization and depolarization. In N.P.Movchan (Ed.), *Physiological mechanisms of main nervous processes*: *Proceeding of Leningrad society of natural scientists*. Leningrad, **75**, № 5, p. 31–40.
Stellar, E. (1954). The physiology of motivation. Psychological Review, **61**, 5-22.
Sufianova, G.Z., Murik, S.E., Sufianov, A.A., Usov, L.A., Shapkin, A.G., Taborov, M.V. (2002). Functional estimation of CPA neuroprotective action by EEG at focal rat brain ischemia. *Bulletin of Eastern-Siberian Scientific Center SB RAMS*, № 6, **1**, p.148–154.
Vasilevskaya, N.E. (1971). *On the function and structure of visceral-chemical analyzer*. Leningrad: Publishing House of Leningrad State University.
Vasiliyev L.L. (1925). About the main functional states of nervous tissue. *New in reflexology and physiology of nervous system*. Leningrad – Moscow, 1–41.
Vvedenskiy, N.E. (1901). *Excitement, inhibition and narcosis*. St. Petersburg.
Winer, N. (1968). *Cybernetics* (1948). M.: Soviet radio.
Zambrzhitskiy, I.A. (1972). *Limbic part of the cerebrum*. – Moscow: Medicina.